\documentclass[11pt]{article}

\usepackage[letterpaper,margin=2cm]{geometry}
\usepackage[T1]{fontenc}

\usepackage{graphicx}
\usepackage{amsmath}
\usepackage{amssymb}
\usepackage[small]{caption}

\usepackage{achemso}

\usepackage{mathptmx}

\newcommand\e[1]{\ensuremath{_{\text{#1}}}}
\newcommand\ex[1]{\ensuremath{^{\text{#1}}}}

\begin{document}

\title{Thermodynamics of guest-induced structural transitions in hybrid
organic--inorganic frameworks}

\author{Fran\c{c}ois-Xavier Coudert$^a$,
Marie Jeffroy$^b$,
Alain H. Fuchs$^c$,\\
Anne Boutin$^b$*,
Caroline Mellot-Draznieks$^a$*}

\date{\today}

\maketitle

\begin{center}

$^a$ Department of Chemistry, University College London, \\
20 Gordon Street, London, WC1H 0AJ, United Kingdom.

\bigskip

$^b$ Laboratoire de Chimie Physique, CNRS and Univ. Paris-Sud, \\
F-91405 Orsay, France.

\bigskip

$^c$ \'{E}cole nationale sup\'{e}rieure de chimie de Paris (Chimie ParisTech) \\
and Univ. Pierre et Marie Curie, F-75005 Paris, France.

\bigskip

* Corresponding authors: 
\texttt{anne.boutin@lcp.u-psud.fr}\par
and
\texttt{c.mellot-draznieks@ucl.ac.uk}

\end{center}

\vskip 10mm

\renewcommand{\baselinestretch}{1.4}

\abstract{
We provide a general thermodynamic framework for the understanding of
guest-induced structural transitions in hybrid organic--inorganic
materials. The method is based on the analysis of experimental adsorption
isotherms. It allows the determination of the free energy differences
between host structures involved in guest-induced transitions, especially
hard to obtain experimentally. We discuss the general case of adsorption
in flexible materials and show how a few key quantities, such as pore
volumes and adsorption affinities, entirely determine the phenomenology
of adsorption, including the occurrence of structural transitions. Based
on adsorption thermodynamics, we then propose a taxonomy of guest-induced
structural phase transitions and the corresponding isotherms. In
particular, we derive generic conditions for observing a double
structural transition upon adsorption, often resulting in a two-step
isotherm. Finally, we show the wide applicability and the robustness of
the model through three case studies of topical hybrid organic--inorganic
frameworks: the hysteretic hydrogen adsorption in
Co(1,4-benzenedipyrazolate), the guest-dependent gate-opening in
Cu(4,4$'$-bipyridine)(2,5-dihydroxybenzoate)$_2$ and the CO$_2$-induced
``breathing'' of hybrid material MIL-53.
}

\pagebreak

\section*{Introduction}

Organic--inorganic framework materials display an extremely large range of
crystal structures and host-guest properties, ranging from coordination
polymers, porous metal--organic frameworks (MOFs), and extended inorganic
hybrids.\cite{Kitagawa_rev1, Rosseinsky_rev, Cheetham2006, Cheetham2008,
Ferey_AccChemRes, Kepert} This makes them an important class of materials
with potentially major impact in adsorption/separation technologies of
strategic gas (H$_2$, CO$_2$ or CH$_4$).\cite{Rowsell, Latroche,
Eddaoudi, Banerjee} The combination of tunable porosity, the
functionalization of the internal surface together with the structural
flexibility of the host opens the way to an extremely rich host-guest
chemistry, putting this class of materials in a unique position.

The distinctive chemistry that results from the flexibility of hybrid
frameworks may be set against that of zeolites, that are characterized by
a relatively limited framework flexibility and a high thermal stability
due to the strength of the metal-oxygen bonds (Si--O bonds are among the
strongest covalent bonds known), allowing their permanent porosity upon
adsorption-desorption processes. By contrast, hybrid materials involve
significantly weaker bonds (coordinative bonds, $\pi$--$\pi$ staking,
hydrogens bonds...) that are responsible for their intrinsic structural
flexibility. Thus, one fascinating aspect of hybrid frameworks is the
ability of a subclass of structures to behave in a remarkable
guest-responsive fashion.\cite{Bradshaw, Rosseinsky, Kitagawa_rev,
Kitagawa_rev2} As recently classified by Kitagawa et
al.,\cite{Kitagawa_rev, Kitagawa_rev2} such hybrid frameworks exhibit a
variety of guest-induced structural phase transitions upon gas adsorption
and desorption. They are typically reported to possess a bistable
behavior controlled by an external stimulus such as gas pressure.

It is striking that a substantial number of experimental adsorption data
collected on guest-responsive hybrid frameworks exhibit S-shape or
step-wise adsorption isotherms, frequently assorted with hysteresis
loops, this for a surprisingly large variety of polar and non polar
sorbates (CO$_2$, CH$_4$, methanol, ethylene...).\cite{Kondo, Noguchi,
Maji, Chen, Shimomura, Matsuda, Goto, Hu, Tanaka} They include a recent
example of hysteretic H$_2$ multi-step isotherms\cite{Long} that do not
conform to any of the IUPAC isotherm types.\cite{IUPAC_isotherms} Despite
the richness of the experimental data at hand on these systems, the
fundamental thermodynamics that underlie these particularly complex
\{host-guest\} systems is far from being understood today and its
phenomenology remains to be developed. In this respect, we may put
forward two strictly distinct situations.

For a number of hybrid frameworks, the S-shape of adsorption isotherms
may be clearly attributed to the very large pore size of the material and
the occurrence of strong sorbate-sorbate interactions reminiscent of the
bulk, as shown in the case of CO$_2$ adsorption in a series of
MOFs.\cite{Millward, Walton} In these situations, the shape of the
isotherm does not arise from a phase transition of the host but from the
details of the structural arrangement of the adsorbed phase, similarly to
what occurs for a number of \{zeolite, guest\} systems (e.g.
\{silicalite-1/heptane\}\cite{Smit} and
\{AlPO$_4$-5/methane\}\cite{Lachet}). The phenomenology of such systems has
been extensively studied and is appropriately described in the frame of
the widely used Grand Canonical ensemble. We will not consider them
further in this paper.

For another subclass of hybrids, which is the primary focus of this
article, the presence of one or more steps in the isotherms has been
attributed to guest-induced structural transitions of the host material.
In this case, the phenomenology of gas adsorption is deemed to be far
more complex, as shown experimentally by the great variety of isotherms
profiles, hysteresis loops and types of phase transitions (classified as
amorphous-to-crystal and crystal-to-crystal
transformations).\cite{Kitagawa_rev} For example, among the most
eye-catching guest-induced phase transitions of hybrids are the so-called
``gate-opening'' and ``breathing'' phenomena, attracting much attention
due to their potential applications in sensing, gas separation or low
pressure gas storage.\cite{Kaneko, Kitaura, Tanaka} ``Gate-opening''
typically involves an abrupt structural transition between a non-porous
state and a porous crystalline host that is induced by gas adsorption. It
is characterized by highly guest-dependent ``gate-opening'' and
``gate-closing'' pressures (on the adsorption and desorption branches
respectively).\cite{Kitaura} The ``breathing'' phenomenon includes two
recent examples, MIL-53\cite{Serre} and MIL-88\cite{MIL88} transition
metal terephtalates, that exhibit massive guest-induced
crystal-to-crystal transformations. While MIL-88 exhibits a gradual cell
expansion up to ~200~\% upon fluid adsorption,\cite{MIL88_2} an abrupt
structural transition is observed in MIL-53 upon H$_2$O\cite{Serre} or
CO$_2$ adsorption,\cite{MIL53_AdvMater} with a transition resulting in a
$\sim$38~\% cell variation.

Despite a very rich experimental corpus of data, there is a dearth of
systematic understanding of the thermodynamics of guest-responsive hybrid
frameworks. Key questions remains to be answered such as, for example:
(i) predicting the occurrence or absence of a guest-induced transition
for a given \{host, gas\} system and predicting the resulting shape of the
adsorption isotherm; (ii) elucidating the thermodynamics that underlie
the guest-induced transition itself, including the relative stabilities
of the two states; or (iii) identifying the key factors that drive the
strong guest-dependence of the gate-opening processes.

One of the reasons that have hindered the systematic understanding of
guest-responsive hybrid frameworks until now is that the thermodynamics
of the two concomitant processes (host-guest interactions and the host
structural transition) cannot be easily deconvoluted experimentally, even
though it is generally agreed that the interplay between the host-guest
interactions and the energetic cost of the structural transition governs
the guest-induced structural transitions. The current approach in the
field relies on the characterisation of the energetics of the adsorption,
typically using calorimetric measurements which however do not
distinguish the host and host-guest contributions. Alternatively,
forcefield-based and DFT single point energy calculations\cite{Coombes,
Ramsahye_JPCC} have been used to elucidate the relative energetic
contributions pertaining to the host and to the guest adsorption,
although leaving the thermodynamic picture incomplete.

A complete description of the thermodynamics of guest-responsive
frameworks can be obtained directly by carrying simulations of the fully
flexible solid in presence of adsorbate, in the so-called osmotic
ensemble. However, this method puts a demanding constraint on the
forcefield development, as they are required to describe the full energy
landscape of the host material, including the structural transition
itself. Moreover, it is frequent with hybrid materials that the low
crystallinity of phases involved in the transition hinders their
structural determination from powder data, therefore eliminating the
possibility of simulating their adsorption isotherms.

The present paper aims at rationalizing the thermodynamics of adsorption
in flexible frameworks when guest-induced structural transitions of the
host are involved. We first develop a thermodynamic description of the
process of adsorption in flexible structures, and then devise a method to
calculate the difference in free energy between the different structures
involved in guest-induced structural transitions. This method relies on
adsorption isotherms, which can be obtained experimentally. We then use
the description developed to discuss the general case of type~I
adsorption in flexible materials, and we show that the thermodynamics of
this process depends only on a few key quantities (pore volumes,
adsorption affinities). This allows us to develop a taxonomy of the
different behaviors that can be observed upon adsorption, and the
resulting types of isotherm. Finally, we show the wide applicability and
the robustness of our free energy calculation method by applying it to
three topical guest-responsive hybrid materials: we study the hysteretic
hydrogen adsorption in Co(1,4-benzenedipyrazolate), the guest-dependent
gate-opening in Cu(4,4$'$-bipyridine)(2,5-dihydroxybenzoate)$_2$ and the
CO$_2$-induced ``breathing'' of hybrid material MIL-53.


\section*{Thermodynamic potential in the osmotic ensemble}

\subsection*{Expression of the thermodynamic potential of host-guest
systems}

We consider here the general process of adsorption of a fluid in a
nanoporous material, where the host framework undergoes structural phase
transitions induced by the adsorption of the fluid. It has been shown
that this process is most appropriately described in the osmotic
statistical ensemble,\cite{Brennan, dePablo_osmotic, jeffroy, Snurr, Shen} where the
control parameters are the number of molecules of the host framework
$N\e{host}$, the chemical potential of the adsorbed fluid $\mu\e{ads}$,
the mechanical constraint $\sigma$ exerted on the system (which is simply
here the external pressure $P$) and the temperature $T$. The osmotic
ensemble in this formulation\cite{dePablo_osmotic} is an extension of the
grand canonical ensemble that accounts for the presence of a flexible
host material with variable unit cell. Its thermodynamic potential
$\Omega\e{os}$ and configurational partition function $Z\e{os}$ are the
following:\cite{Faure} \begin{equation} \Omega\e{os} = U - TS -
\mu\e{ads}N\e{ads} + PV \end{equation} \begin{equation} Z\e{os} = \sum_V
\sum_{N\e{ads}} \sum_{\mathbf{q}} \exp\left[ -\beta U(\mathbf{q}) +
\beta\mu\e{ads}N\e{ads} - \beta PV\right] \end{equation} where
$\mathbf{q}$ denotes the positions of the atoms of the system (host and
adsorbate) and $\beta=1/kT$ ($k$ being the Boltzmann constant).

In order to study the relative stability of different phases of the host
structure upon adsorption, our method relies on the factorization of the
expressions above in two contributions, one characterizing the framework
structures themselves (i.e. their free energy), the other describing the
fluid adsorption in each solid phase involved in the transition. The
summation over the positions $\mathbf{q}$ of atoms in the system is thus
split into a double summation over the coordinates of the material,
$\mathbf{q}\e{host}$, and coordinates of the atoms of the adsorbate,
$\mathbf{q}\e{ads}$. The total energy is the sum of the energy of the
host, the energy of the adsorbed fluid and the host-adsorbate
interactions: \begin{equation} U(\mathbf{q}) =
U\e{host}(\mathbf{q}\e{host}) + U\e{ads}(\mathbf{q}\e{ads}) +
U\e{host-ads}(\mathbf{q}\e{ads}, \mathbf{q}\e{host}) \end{equation} As
our interest lies in the determination of adsorption-induced structural
transitions of the host material, we make the assumption that while the
fluid adsorption may favor one phase or another, each single structure is
only marginally changed by the adsorption of fluid. This is translated
into a mean field approximation by writing that the host-adsorbate
interactions depend only on the average positions of the atoms (and unit
cell vectors) of the material in a given phase~$i$: \(
U\e{host-ads}(\mathbf{q}\e{ads}, \mathbf{q}\e{host}) \simeq
U\e{host-ads}(\mathbf{q}\e{ads}; \left<\mathbf{q}\e{host}\right>_i) \).
Using this separation of variables, the configurational partition
function can be written as a sum of configurational partition functions
for each phase~$i$ of the solid: $Z\e{os} = \sum_i Z\e{os}\ex{($i$)}$.
The configurational partition function of any given phase~$i$ of the
material is then given by: \begin{equation} \begin{array}{r@{\ }c@{\ }l}
Z\e{os}\ex{($i$)} & = & \displaystyle \left(\sum_{V\in i}
\sum_{\mathbf{q}\e{host}\in i} \exp\left[ -\beta
U\e{host}(\mathbf{q}\e{host}) - \beta PV \right] \right) \\ &&
\displaystyle \times \left( \sum_{N\e{ads}} \sum_{\mathbf{q}\e{ads}}
\exp\left[ -\beta U\e{ads}(\mathbf{q}\e{ads}) - \beta
U\e{host-ads}(\mathbf{q}\e{ads}; \left<\mathbf{q}\e{host}\right>_i) +
\beta\mu\e{ads}N\e{ads} \right] \right) \end{array} \end{equation} It is
apparent that the first term is the configurational partition function
(restricted to phase~$i$) of the isolated host structure in the
$(N\e{host},P,T)$ ensemble, $Z\e{host}\ex{($i$)}$. The second term can be
recognized as the grand canonical configurational partition function
$Z\e{GC}\ex{($i$)}$ of the fluid in the external field created by the
host framework considered as rigid. Thus, for each host phase~$(i)$,
$Z\e{os}\ex{($i$)} = Z\e{host}\ex{($i$)} \times Z\e{GC}\ex{($i$)}$, can
be written in terms of the thermodynamic potentials for each ensemble:
$\Omega\e{os}\ex{($i$)} = G\e{host}\ex{($i$)} + \Omega\ex{($i$)}$, where
$\Omega$ is the grand canonical potential for the adsorbate and
$G\e{host}$ is the free enthalpy of the host material.

To express the grand canonical potential as a function of quantities
directly available from experiments or simulations, we calculate it from
its derivatives, using the fundamental relation involving the chemical
potential $\mu\e{ads}$:\cite{Peterson, Puibasset, Cailliez}
\begin{equation}
\left(\frac{\partial\Omega}{\partial\mu\e{ads}}\right)_{V,T} = -N\e{ads}
\end{equation} Integrating this equation and taking into account that
$\Omega(P=0)=0$, the grand canonical potential $\Omega$ can be rewritten
as a function of pressure~$P$ rather than chemical potential
$\mu\e{ads}$, as follows: \begin{equation} \Omega(P) = -\int_0^P
N\e{ads}(p) \left(\frac{\partial\mu}{\partial p}\right)_{T,N} \mathrm{d}p
\end{equation} which can be simplified further by introducing the molar
volume of the pure fluid, \( V\e{m} = (\partial\mu/\partial P)_{T,N} \):
\begin{equation} \Omega(P) = -\int_0^P N\e{ads}(p) V\e{m}(p) \mathrm{d}p
\end{equation} The free enthalpy of the host can also be written as a
function of the free energy of the host at zero pressure, $F\e{host}$:
$G\e{host} = G\e{host}(P=0) + PV = F\e{host} + PV$.

Finally, this leads to a complete expression of the osmotic potential
which involves only three key parameters: the free energy of the solid
phase, the adsorption isotherm of fluid inside that phase
($N\e{ads}(T,P)$) and the molar volume of the pure fluid as a function of
pressure: \begin{equation}\label{eq:Omega_os} \Omega\e{os}(T,P) =
F\e{host}(T) + PV - \int_0^P N\e{ads}(T,p) V\e{m}(T,p) \mathrm{d}p
\end{equation}


\subsection*{Prediction of the free energy of phases involved in
structural transitions}

We now have an expression for the thermodynamic osmotic potential as a
function of gas pressure for each solid phase. The comparison of
$\Omega\e{os}$ for each phase of the host allows us to determine the
relative stability of the phases, the number of structural transitions
that will take place and the pressure at which they occur. This requires
the prior determination of the three quantities in
Equation~\ref{eq:Omega_os}. The relative free energies of the empty solid
phases ($F\e{host}\ex{($i$)}$) are especially difficult to evaluate both
experimentally and by simulation methods. While relative energies between
solid phases can be experimentally measured, for example by differential
scanning calorimetry or the measurement of dissolution
enthalpies,\cite{Navrotsky} or calculated using first-principles methods,
the determination of relative \emph{free energies} is a much harder task.

By contrast, the pressures at which guest-induced transitions happen are
relatively easy to measure experimentally, either because they show up as
steps on the adsorption isotherm, or by calorimetry (structural
transitions often result in a sudden change in heat of adsorption) or
even by \emph{in situ} X-ray diffraction studies at different gas
loadings. For that reason, we propose here a method to calculate the
relative free energies of the empty solid phases ($F\e{host}\ex{($i$)}$)
from the readily available experimental quantities that are the sorption
isotherms and the phase transition pressure, using
Equation~\ref{eq:Omega_os}.

\begin{figure} \begin{center}
\includegraphics[width=7.0cm]{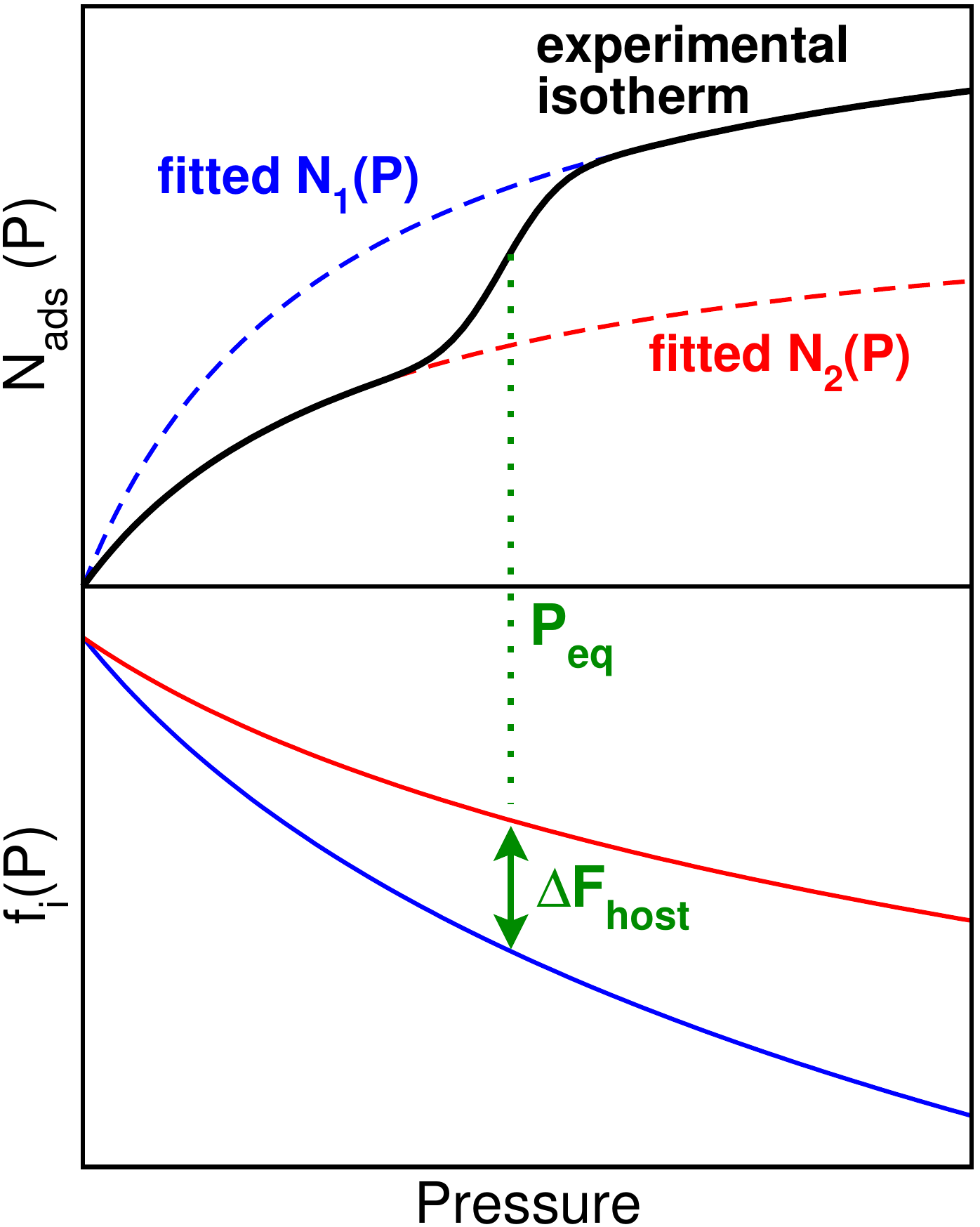}
\caption{\label{fig:method}Schematic representation of the
determination of free energy difference $\Delta F\e{host}$ from an
experimental adsorption isotherm (upper panel) by calculation of
$f_i(P)$, the pressure-dependent part of the thermodynamic potential of
the osmotic ensemble (lower panel).}
\end{center} \end{figure}

The quantities involved in Equation~\ref{eq:Omega_os} and that we need to
calculate free energy differences are now the transition pressures, the
adsorption isotherms for each isolated phase ($N\e{ads}\ex{($i$)}(P)$),
the molar volume of the pure adsorbate as a function of pressure
($V\e{m}(P)$) and the unit cell volume of each phase ($V_i$). That last
two are rather straightforward to obtain: $V_i$ is known from crystal
structures determined by X-ray diffraction and $V\e{m}(P)$ can be found
tabulated for a large number of gases or alternatively approximated by
the relation for an ideal gas, $V\e{m}(P) = RT/P$. The pressures at which
structural transitions occur may be determined from the experimental
adsorption and desorption isotherms of the material or other methods
described above. However, the ``virtual'' rigid-host isotherms
$N\e{ads}\ex{($i$)}(P)$ related to each phase are harder to obtain
experimentally: they correspond to each of the solid phases assuming the
absence of transition over the whole range of pressures, so that only
parts of these rigid-host isotherms are observed experimentally, as shown
in Figure~\ref{fig:method}. To overcome this issue, we propose here to
fit the distinct parts of a stepped isotherm to obtain full
``rigid-host'' isotherms needed for each of the phases. They correspond
to the dashed lines in Figure~\ref{fig:method}. Alternatively, it is
possible to use isotherms calculated from Grand-Canonical Monte Carlo
simulations in each phase, with the host structure being considered
rigid.\cite{jeffroy}

Let us illustrate here the method on an isotherm possessing a single step
that corresponds to a guest-induced structural transition
(Figure~\ref{fig:method}) between two phases labeled \textbf{1} and
\textbf{2}, assuming \textbf{1} and \textbf{2} have been structurally
characterized and have known unit cell volumes, $V_1$ and $V_2$. From the
step in the experimental isotherm, the pressure $P\e{eq}$ of the
structural transition can be determined and the two parts of the isotherm
(below and above the step) are fitted by Langmuir
isotherms\cite{note_fit_type} (dashed lines on Figure~\ref{fig:method}).
From the fitted isotherms $N\e{ads}\ex{($i$)}(P)$, we can now plot
(Fig.~\ref{fig:method}, lower panel) the two functions $f_1(P)$ and
$f_2(P)$ defined as: \begin{equation} \label{eq:fi} f_i(P) =
\Omega\e{os}\ex{($i$)}(P) - F\e{host}\ex{($i$)} = PV_i - \int_0^{P}
N\e{ads}\ex{($i$)}(p) V\e{m}(p) \mathrm{d}p \end{equation} At the
transition pressure, the thermodynamic equilibrium between the two phases
$\Omega\e{os}\ex{(1)}(P\e{eq}) = \Omega\e{os}\ex{(2)}(P\e{eq})$ and thus
$\Delta F\e{host} = f_1(P\e{eq}) - f_2(P\e{eq})$.

It is clear that this method is straightforward to extend to cases where
more than two solid phases come into play, simply by applying the above
construction to each structural transition. It is also worth noting that,
if sorption isotherms are available for different temperatures, both
energy and entropy differences, $\Delta U\e{host}$ and $\Delta
S\e{host}$, can be extracted from the free energies: $\Delta
F\e{host}(T) = \Delta U\e{host} - T \Delta S\e{host}$.

\section*{Taxonomy of guest-induced transitions in flexible frameworks}

In this section, we apply the equations described above to the case of a
nanoporous, flexible material that has two different metastable phases,
and for which fluid adsorption follows type~I isotherms in each
structure. This case, including the possibility of one structure having
no microporosity at all, has frequently been observed for hybrid
organic--inorganic frameworks. We will show that this case yields a
straightforward analytical expression for $\Delta\Omega\e{os}(P)$ and
enables us to propose a phenomenology of guest-induced structural
transitions induced by adsorption and a classification of the resulting
isotherms into distinct categories.

Let us consider that, for each phase~$i$ of the material, the adsorption
happens in the gas phase (considered ideal) and follows a type~I
isotherm.\cite{IUPAC_isotherms} In order to keep the analytical
expressions simple and to include only a few key physical quantities, the
Langmuir equation is used to describe the gas adsorption:
\begin{equation}\label{eq:Langmuir} N\e{ads}\ex{($i$)} =
\frac{K_iP}{1+\frac{K_iP}{N\e{max}^i}} \end{equation} where $K_i$ is the
Henry constant for adsorption, which measures the adsorption affinity,
and $N\e{max}^i$ is the number of adsorbed gas molecules at the plateau
of the isotherm. Plugging Equation~\ref{eq:Langmuir} into
Equation~\ref{eq:Omega_os}, and taking for $V\e{m}$ the expression of the
ideal gas, the osmotic potential of phase~$i$ can be written as follows:
\begin{equation} \Omega\e{os}\ex{($i$)} = F\e{host}\ex{($i$)} + PV_i -
\int_0^P \frac{K_i}{1+\frac{K_i P}{N\e{max}^i}} \mathrm{d}P =
F\e{host}\ex{($i$)} + PV_i - N\e{max}^i RT \ln\left( 1+\frac{K_i
P}{N\e{max}^i} \right) \end{equation}

The guest-induced structural transitions of the host material are
dictated by the osmotic potentials $\Omega\e{os}\ex{($i$)}$ of the
different solid phases as a function of pressure. We discuss in this
section the conditions of occurrence of guest-induced transitions. We
consider two structures of a single host material, labeled~\textbf{1}
and~\textbf{2} in such a way that, in the absence of adsorbate,
structure~\textbf{1} is more stable than the structure~\textbf{2} (i.e.,
$\Delta F\e{host} = F\e{host}\ex{(2)} - F\e{host}\ex{(1)}$ is positive).
The osmotic potential difference between these structures, as a function
of pressure, is expressed as: \begin{equation}
\label{eq:diff_omega_langmuir_withvol} \Delta\Omega\e{os} = \Delta
F\e{host} + P\Delta V - RT \left[ N\e{max}\ex{(2)} \ln\left(1+\frac{K_2
P}{N\e{max}\ex{(2)}} \right) - N\e{max}\ex{(1)} \ln\left(1+\frac{K_1
P}{N\e{max}\ex{(1)}} \right) \right] \end{equation} All the terms in this
equation have a clear physical meaning and the full thermodynamic
behavior can be discussed from there. To shorten the discussion, we now
proceed to simplifying Equation~\ref{eq:diff_omega_langmuir_withvol}. In
all the cases presented in this article, gas adsorption happens in a
range of pressures for which the term $P\Delta V$ in the expression above
is of limited importance. To simplify the analytical formulas and the
discussion that follows, $P\Delta V$ will thus be neglected it in the
rest of this section. Moreover, we replace the saturation values of the
isotherms, $N\e{max}^i$, with the accessible porous volume of the
material in phase~$i$, $V\e{p}\ex{($i$)}$, by writing $N\e{max}^i = \rho
V\e{p}\ex{($i$)}$, with $\rho$ the density of the adsorbed gas at high
pressure.\cite{note_density}

Equation~\ref{eq:diff_omega_langmuir_withvol} can then be rewritten as
follows: \begin{equation} \label{eq:diff_omega_langmuir}
\Delta\Omega\e{os}(P) = \Delta F\e{host} - RT\rho \left[ V\e{p}\ex{(2)}
\ln\left(1+\frac{K_2 P}{\rho V\e{p}\ex{(2)}} \right) - V\e{p}\ex{(1)}
\ln\left(1+\frac{K_1 P}{\rho V\e{p}\ex{(1)}} \right) \right]
\end{equation}

We now study the evolution of $\Delta\Omega\e{os}(P)$ given by
Equation~\ref{eq:diff_omega_langmuir} and in particular the solutions of
the equation $\Delta\Omega\e{os}(P) = 0$ (i.e., the structural
transitions). As demonstrated in Appendix~A, the behavior of the system
and the occurrence of guest-induced phase transitions are entirely
determined by five key parameters characteristic of the system: the
difference in free energy between the empty host structures, $\Delta
F\e{host}$, the pore volumes, $V\e{p}\ex{(1)}$ and $V\e{p}\ex{(2)}$, and
the Henry constants, $K_1$ and $K_2$. We can identify four distinct
cases, synthesized in Figure~\ref{fig:iso_types}:

\begin{figure} \begin{center}
\includegraphics[width=12cm]{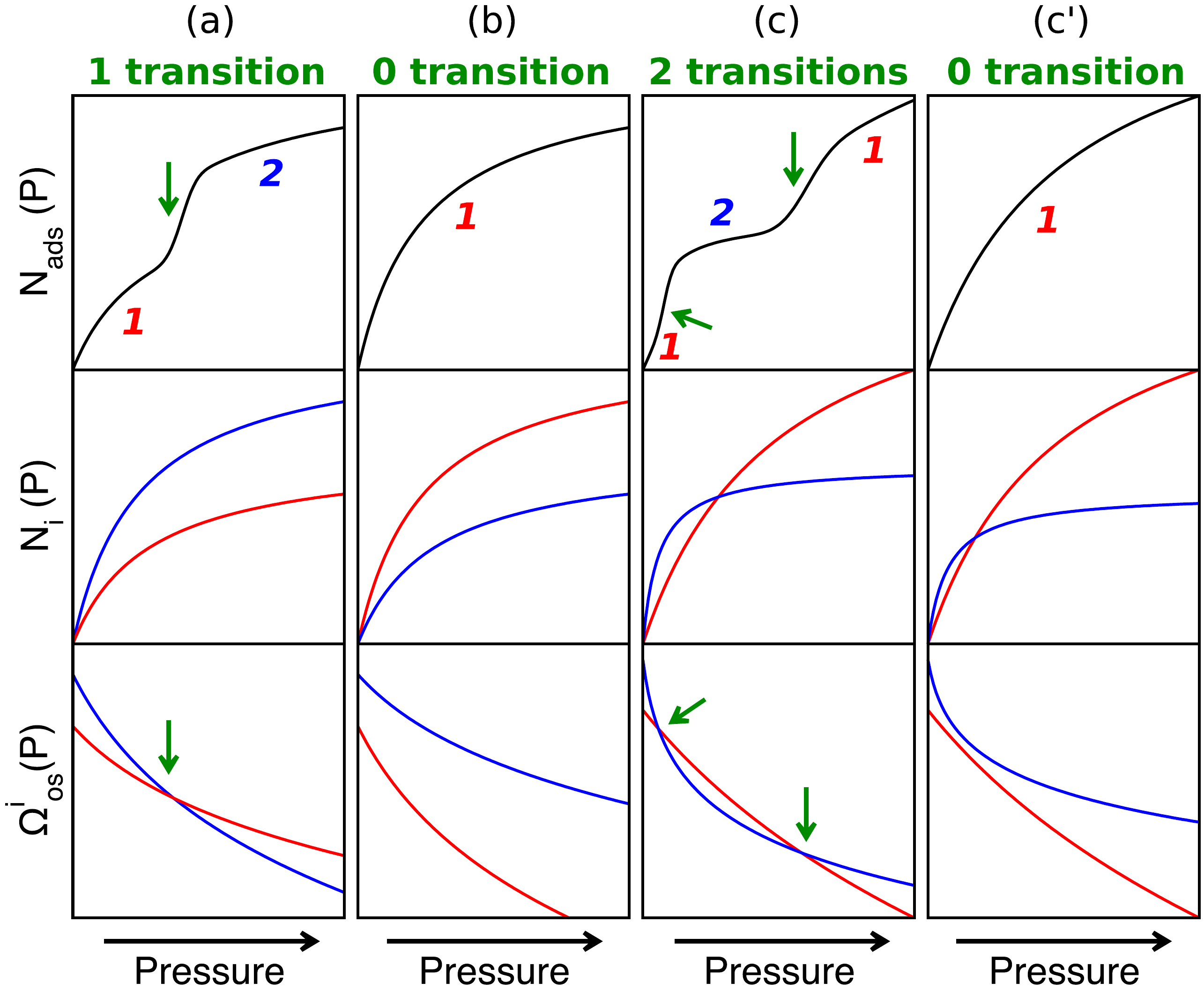}
\caption{\label{fig:iso_types}The four possible cases of
Langmuir-type adsorption in materials with two metastable
phases~\textbf{1} and~\textbf{2}. Top panels show the adsorption
isotherms, middle panels depict the step-free isotherms corresponding to
each phase, and bottom panels plot the osmotic potential of both phases.
Green arrows indicate guest-induced structural transitions.}
\end{center} \end{figure}

\begin{itemize}

\item \textbf{\boldmath Case a, $V\e{p}\ex{(2)} > V\e{p}\ex{(1)}$:} Under
this condition, whatever the values of $K_1$, $K_2$ and $\Delta
F\e{host}$, only one structural transition is observed.
Structure~\textbf{1} will be more favorable at low pressure, and as
pressure increases, the larger accessible pore volume of
structure~\textbf{2} will make it more and more favorable, leading to a
structural transition upon adsorption.\cite{note_case_a} This transition
leads to a one-step isotherm (see Fig.~\ref{fig:iso_types},
column~\textbf{a}). This case is quite common and is actually observed in
many of the so-called ``third generation coordination
polymers''\cite{Kitagawa_rev} that have attracted a lot of interest in
the recent years. In the next section, we will show that our method can
usefully be applied to two materials exhibiting crystal-to-crystal
transformations of this type, including the example of
\hbox{[Cu(4,4$'$-bipy)(dhbc)$_2$]}, and can apply to ``gating
processes'',\cite{Kitaura} where one of the structures has no
microporosity at all.

\item \textbf{\boldmath Case b, $V\e{p}\ex{(1)} > V\e{p}\ex{(2)}$ and
$K_1 > K_2$:} Under these conditions, we can predict that no structural
transition will be observed. This stems from the fact that all factors
favor structure~\textbf{1}: the empty structure has lower free energy, it
has higher adsorption affinity and higher pore volume
(Fig.~\ref{fig:iso_types}, column~\textbf{b}).

\item \textbf{\boldmath Cases c and c$'$, $V\e{p}\ex{(1)} >
V\e{p}\ex{(2)}$ and $K_2 > K_1$:} Under these conditions, two different
behaviors must be distinguished. At low pressure, structure~\textbf{1} is
favored. At high pressure, it is also favored because it has a larger
pore volume. In between, however, there might exist a regime where
structure~\textbf{2} is thermodynamically favored because of its larger
adsorption affinity. Whether this regime occurs or not depends on the
balance between the terms in Equation~\ref{eq:diff_omega_langmuir}. If it
happens, the system will undergo two successive structural transitions
upon adsorption (from phase~\textbf{1} to phase~\textbf{2}, then back to
phase~\textbf{1}).

The condition for the occurrence of the double transition can be most
easily expressed as an upper bound on the value of $\Delta F\e{host}$:
\begin{equation} \frac{\Delta F\e{host}}{\rho RT} < \left(V_2-V_1\right)
\ln\left( \frac{K_2V_1 - K_1V_2}{V_1-V_2} \right) + V_1\ln K_2 - V_2\ln
K_1 \end{equation} Thus, if $\Delta F\e{host}$ is small enough (or, seen
the other way, if $\Delta K$ is large enough), there will be a domain of
stability for structure~\textbf{2}, and two transitions will be observed
upon adsorption. This leads to an adsorption isotherm with two successive
steps, as shown in third column of Fig.~\ref{fig:iso_types} (\textbf{case
c}). On the other hand, if $\Delta F\e{host}$ is too large (or $\Delta K$
too small), there will be no domain of stability for
structure~\textbf{2}, and structure~\textbf{1} will be the most stable
phase in the entire pressure range (see last column of
Fig.~\ref{fig:iso_types}, \textbf{\boldmath case c$'$}), leading to the
absence of structural transition. We will show later that the occurrence
of a double guest-induced structural transition was indeed suggested
experimentally in the case of CO$_2$ adsorption in
MIL-53.\cite{MIL53_JACS,MIL53_AdvMater}

\end{itemize}

The cases considered above correspond to guest-induced transitions
between two distinct host structures. More generally, this taxonomy can
be extended to more complex cases where more than two host structures are
involved.


\section*{Case studies of guest-induced transitions of hybrid frameworks}

In this section, the above method is used to investigate the
thermodynamics of three distinct cases of interest. The first one relates
to the study of a recently discovered H$_2$-induced phase transition in a
cobalt-based hybrid material. The second aims at studying the topical
case of gate-opening processes. Finally, we will show how our method may
be valuably applied to elucidate the thermodynamics of the rather complex
case of the ``breathing'' of MOFs upon gas adsorption.

\subsection*{Structural phase transition in
Co(1,4-benzenedipyrazolate) induced by {\boldmath H$_2$} adsorption}

An interesting H$_2$ stepped isotherm was reported recently by Choi
\emph{et al.}, in the case of the Co(BDP) (BDP =
1,4-benzenedipyrazolate).\cite{Long} The area of hydrogen storage in
hybrid materials is a particularly challenging and topical
one,\cite{Rowsell, Collins} and any improvements in terms of H$_2$
storage capacity and performances are considered highly valuable. Until
now, most hybrid frameworks had exhibited a traditional type~I reversible
H$_2$ adsorption isotherm, that allowed a ranking of the performances of
materials on the sole basis of the weight adsorption capacity. By
contrast, in the case of Co(BDP), the particular eye-catching feature is
a stepped isotherm with a wide hysteresis,\cite{Long} allowing the
adsorption of H$_2$ at high pressure and its storage at lower
pressure,\cite{Zhao, Omary} a highly desirable feature for practical
applications such as transportation. The structural transition in Co(BDP)
is reported to be induced by H$_2$ adsorption and therefore highlights
that subtle energetic and thermodynamic effects are at play, having in
mind that H$_2$ is associated with weak host-guest interactions in hybrid
materials. This case caught our attention in the context of this work.

\begin{figure} \begin{center}
\includegraphics[width=15cm]{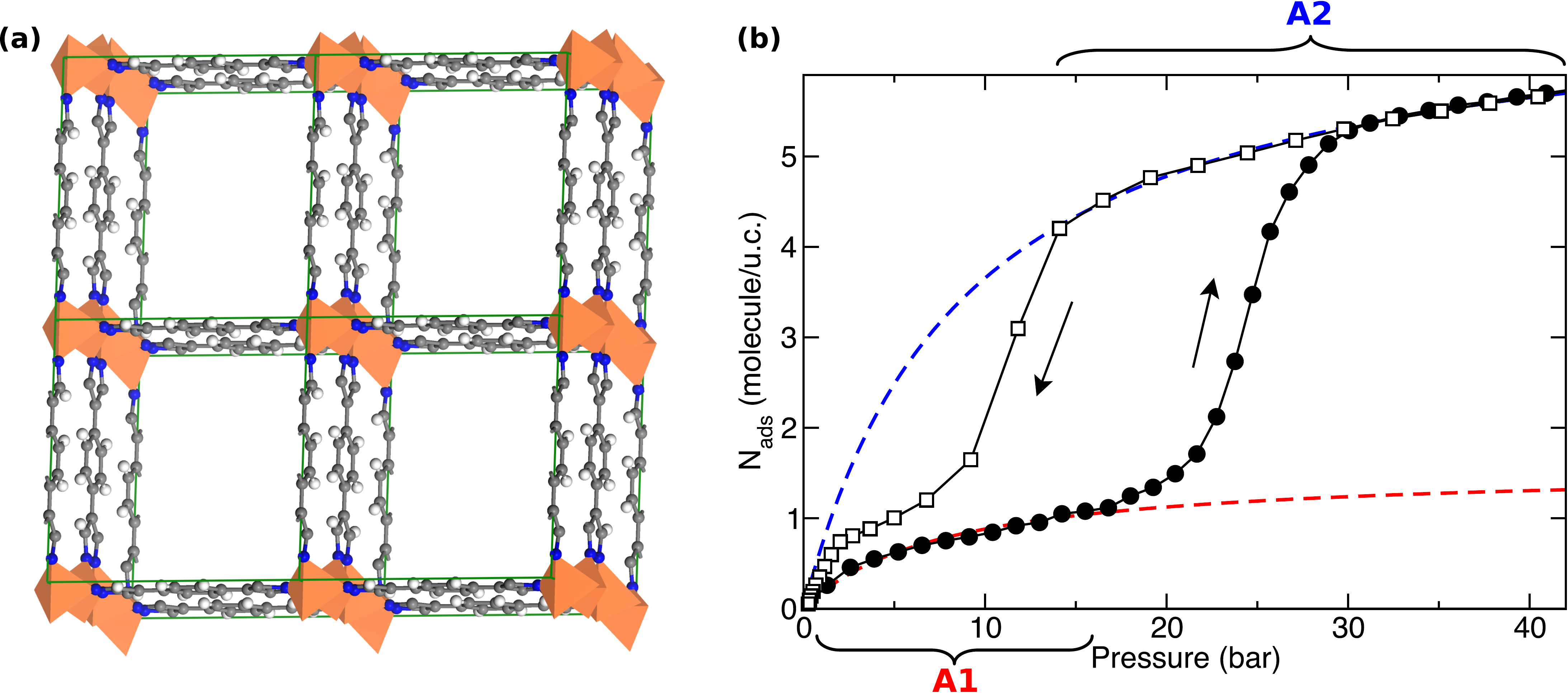}
\caption{\label{fig:Long_iso}%
\textbf{(a)} $2\times 2\times 1$ supercell of structure~\textbf{As},
Co(BDP)$\,\cdot\,$2$\,$DEF$\,\cdot\,$H$_2$O, viewed along the $c$~axis.
\textbf{A2} is believed to have the same framework structure as
\textbf{As}.
\textbf{(b)} Adsorption and desorption isotherms, respectively in filled
circles and empty squares, of H$_2$ in Co(BDP) at 77~K.\cite{Long} The
blue dashed line is the fit of the upper part ($P\ge 15$~bar) of the
desorption isotherm. The red dashed line is the fit of the lower part
($P\le 15$~bar) of the adsorption isotherm.%
}\end{center} \end{figure}

The as-synthesized material Co(BDP)$\cdot$2DEF$\cdot$H$_2$O (DEF =
N,N-di\-ethyl\-formamide), noted hereafter \textbf{As}, has a quadratic
structure presenting \hbox{(10~\AA)$^2$} square channels connected by
narrow openings (see Figure~\ref{fig:Long_iso}a). Its full desolvation
leads to the formation of crystalline Co(BDP), hereafter named
\textbf{A1}, whose structure has not been solved; this process is
reversible. N$_2$ adsorption in Co(BDP) at 77~K results in a multistep
adsorption, interpreted by guest-induced transitions between \textbf{A1},
which has a rather small pore volume, and a fully open framework
\textbf{A2}, believed to have the same framework structure as \textbf{As}
on the basis of similarity of pore volume.

Experimental adsorption and desorption isotherms of H$_2$ in Co(BDP) were
reported at 50~K, 65~K, 77~K and 87~K, showing various complex
multistepped isotherms.\cite{Long} In this study, we have selected the
isotherm at 77~K (Fig.~\ref{fig:Long_iso}b) because of its well-defined
single-step isotherm, which was interpreted as resulting from a
\textbf{A1}$\rightarrow$\textbf{A2} structural transition. We apply here
the method presented above to this experimental isotherm in order to
calculate the free energy difference between \textbf{A1} and \textbf{A2}.
The adsorption isotherm is fitted to a Langmuir equation in the 0--15~bar
region to provide the ``rigid-host'' isotherm of phase \textbf{A1}; the
desorption isotherm is fitted similarly in the 15--40~bar region to
obtain the ``rigid-host'' isotherm of phase \textbf{A2}. Using the
experimental isotherm, we estimate the thermodynamic transition to occur
at $P\e{eq} \approx 15$~bar, considering that the isotherm deviates from
the Langmuir fits at 14~bar and 17~bar, for the desorption and adsorption
branches respectively. Applying Equation~\ref{eq:fi} and neglecting the
$PV_i$ terms,\cite{note_pv} we find that the difference in free energy
between \textbf{A1} and \textbf{A2} is of 3.3~kJ/mol ($\pm 0.2$~kJ/mol).
This result is in very good agreement and much more accurate than the
estimation put forth by Choi \emph{et al.} that the energy of the
structure change process lies in the 2--8~kJ/mol range.

It is interesting to note that this value of 3.3~kJ/mol, associated with
an isotherm measured at 77~K, is significantly larger than the thermal
energy $kT$ ($\sim$ 5 times larger). Following the reasoning of Choi
\emph{et al.}, we can then use the calculated free energy difference
between host phases to estimate the heat of adsorption of H$_2$ in
Co(BDP), $\Delta H\e{ads}$, as $\Delta H\e{ads} = \Delta H\e{host} +
\Delta\e{f}H$. The latter term, $\Delta\e{f}H$, is the formation enthalpy
of a \hbox{Co(BDP):H$_2$} clathrate complex and was estimated by Choi
\emph{et al.} to be 3.2~kJ/mol ($\pm 0.3$~kJ/mol). Neglecting entropic
effects at 77~K, this equation allows us to propose a heat of adsorption
of $\sim$6.5~kJ/mol for H$_2$ in Co(BDP). This heat of adsorption is
rather in the lower region of the range typically observed, from 5~to
11~kJ/mol.\cite{Collins}


\subsection*{Gate-opening transition in a flexible coordination polymer}

\begin{figure} \begin{center}
\includegraphics[width=15cm]{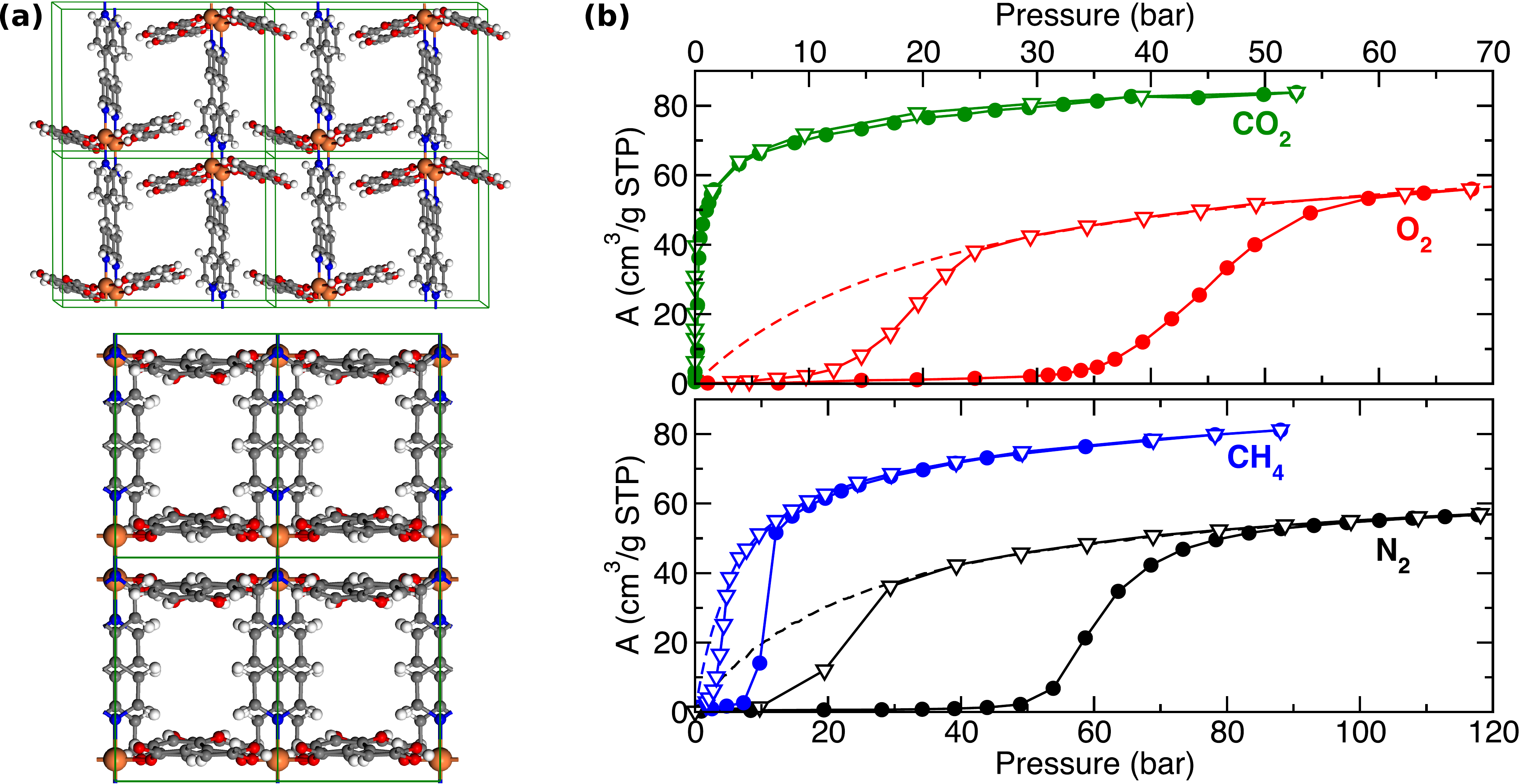}
\caption{\label{fig:Kitaura_iso}%
\textbf{(a)} $2\times 2$ supercell of coordination polymer
\hbox{[Cu(4,4$'$-bipy)(dhbc)$_2$]$\cdot$ H$_2$O}: top, along the
$a$~axis; bottom, along the $c$~axis.
\textbf{(b)} Adsorption and desorption isotherms, respectively in filled
circles and empty triangles, of various small molecules in
\hbox{Cu(4,4$'$-bipy)(dhbc)$_2$} at 298~K, as found in
Ref.~\citenum{Kitaura}: CO$_2$ (green, upper panel), O$_2$ (red, upper
panel), CH$_4$ (blue, lower panel) and N$_2$ (black, lower panel). The
dashed lines are the fits of the upper part of each desorption isotherm
(at pressure higher than the gate-closing pressure) by a Langmuir
equation.%
} \end{center} \end{figure}

Gate-opening in flexible hybrid frameworks occurs when a material exhibit
a structural transition from non-porous to porous structure at a specific
pressure, and has been reported in a number of compounds.\cite{Kaneko,
Kitaura, Tanaka, Rosseinsky} These materials are expected to find
applications as sensors or switches, as well as gas separation. We focus
here on the case of a particular flexible coordination polymer,
\hbox{Cu(4,4$'$-bipy)(dhbc)$_2$$\cdot$ H$_2$O} (4,4$'$-bipy =
4,4$'$-bipyridine; dhbc = 2,5-dihydroxybenzoate), whose as-synthesized
structure~\textbf{Bs} has been solved and which is known to exhibit a
guest-induced structural transition upon adsorption of a large variety of
gases (CO$_2$, O$_2$, CH$_4$ and N$_2$) at 298~K.\cite{Kitaura} The
crystal-to-crystal structural transition occurs between two phases
hereafter labeled~\textbf{B1} and~\textbf{B2}, whose structures have not
been solved. \textbf{B1}, the dehydrated material, shows no microporosity
and, upon gas adsorption, its channels open up at a given
(guest-dependent) gate-opening pressure to yield the fully open
\textbf{B2}. The framework structure of \textbf{B2} is the same as that
of \textbf{Bs}, and is shown in Fig.~\ref{fig:Kitaura_iso}a. It is
composed of interdigitated two-dimension motifs
(Fig.~\ref{fig:Kitaura_iso}a, top) formed by copper~(II) ions linked by
bipyridine and dihydroxybenzoate linkers, and stacked due to $\pi$--$\pi$
interactions between parallel dhbc ligands. Their interdigitation creates
unidimensional channels along the $a$~axis, with a {8~\AA} diameter.

The adsorption isotherms of various small molecules (N$_2$, CH$_4$ and
O$_2$; see Figure~\ref{fig:Kitaura_iso}b) in structure~\textbf{B1} at
298~K present common features: little to no adsorption in the lower
pressure region, followed by a abrupt increase attributed to the
\textbf{B1}$\rightarrow$\textbf{B2} transition. Following
Kaneko\cite{Kaneko} and Kitagawa\cite{Kitaura}, we call here
``gate-opening'' pressure the point of the isotherm at which the
structural transition happens during adsorption. The desorption
isotherms, reversely, show an abrupt drop starting at a pressure that we
will call the ``gate-closing'' pressure, where the
\textbf{B2}$\rightarrow$\textbf{B1} transition takes place. Each
gate-closing pressure is lower than the corresponding gate-opening
pressure, and the isotherms all exhibit hystereses. Only the isotherm for
CO$_2$ is different in that transition happens at such a low pressure
that no hysteresis was detected with the experimental setup.

We apply our method to each set of experimental isotherm (including
CO$_2$) and calculate the difference in free energy between the two
phases \textbf{B1} and \textbf{B2}. As structure~\textbf{B1} is not
porous, only the adsorption isotherm of structure~\textbf{B2} is needed
for the calculation. We use a Langmuir-type fit of the experimental
desorption isotherms in the region of pressure higher than the
gate-closing pressure. The fits, shown as dashed line for each isotherm
in Fig.~\ref{fig:Kitaura_iso}b, are all quite satisfactory. However,
because the pressure of the thermodynamic structural transition can only
be bracketed by the gate-opening and gate-closing pressures, the method
does not lead to a single value of $\Delta F\e{host}$ but to a range of
free energy difference between structures. 

Table~\ref{tab:free_en_coordpolymer} presents the values of free energy
obtained from the isotherms of each gas. In each case, we find that the
nonporous structure is more stable than the open one by 4~kJ/mol ($\pm
0.5$~kJ/mol) at 298~K. It is striking that, using different gas
adsorption isotherms exhibiting different gate-opening and gate-closing
pressure ranges, we obtain such narrow and consistent ranges for the
value of $\Delta F\e{host}$. This excellent agreement between all the
values independently obtained validates our approach and is a strong sign
of the robustness and reliability of the method presented in this
article. Indeed, in the absence of structural characterization of the
transition, we show that a unique mechanism is at play behind the
seemingly different features of all four isotherms.

\begin{table} \begin{center} \small
\begin{tabular}{cccc}\hline
Adsorbate & Gate-opening & Gate-closing
 & Calculated $\Delta F\e{host}$ \\\hline
N$_2$  & 30~bar & 49~bar & 3.3 -- 4.5~kJ/mol \\
CH$_4$ & 7~bar  & 12~bar & 3.6 -- 5.1~kJ/mol \\
O$_2$  & 25~bar & 37~bar & 3.4 -- 4.3~kJ/mol \\
CO$_2$ & $< 2$~bar & $< 2$~bar & $< 6$~kJ/mol \\\hline
\end{tabular}
\caption{\label{tab:free_en_coordpolymer}Gate-opening and gate-closing
pressures extracted from adsorption and desorption isotherms of various
molecules in \hbox{[Cu(4,4$'$-bipy)(dhbc)$_2$]$\cdot$ H$_2$O}, as well as
free energy difference (at 298~K) between the open and closed structures
of the material calculated from each isotherm; see text for details.}
\end{center} \end{table}


\subsection*{``Breathing'' phenomenon in the 3D hybrid framework MIL-53}

\begin{figure} \begin{center}
\includegraphics[width=15cm]{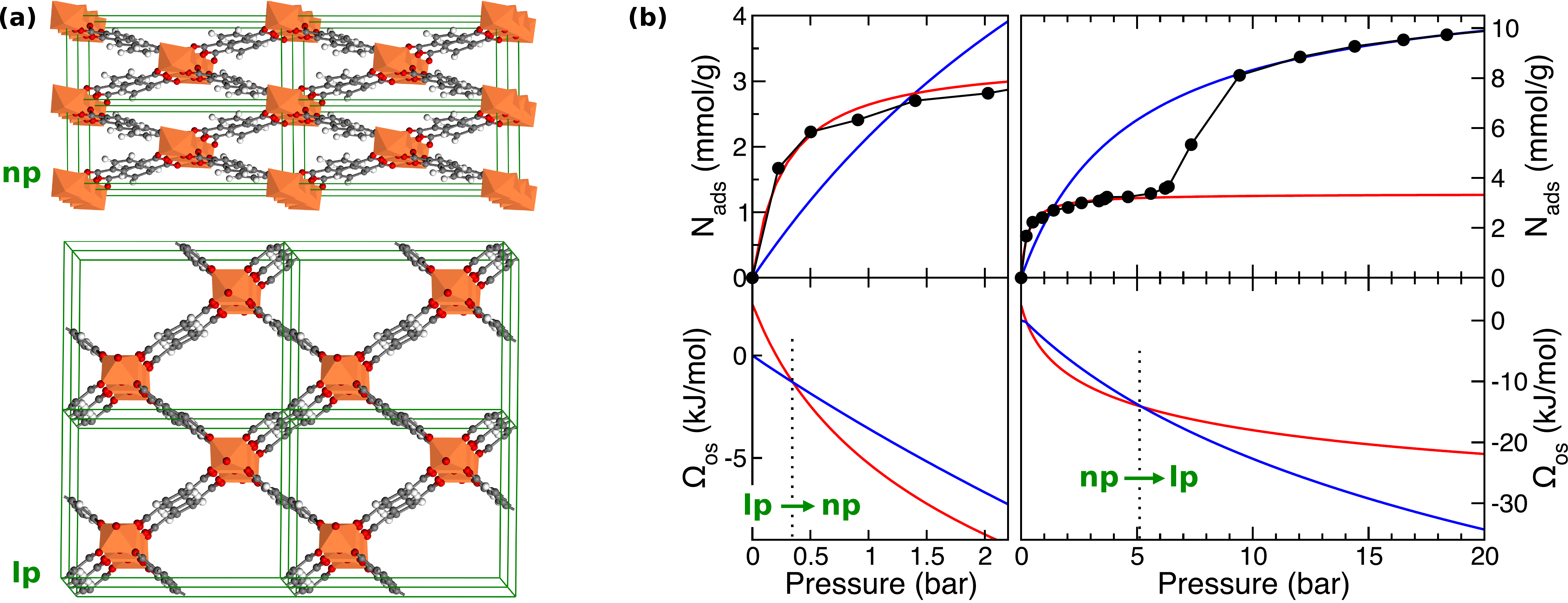}
\caption{\label{fig:iso_CO2_MIL53}%
\textbf{(a)} Narrow-pore (\textbf{np}) and large-pore (\textbf{lp})
forms of the MIL-53 structure, viewed along the axis of the
unidimensional channels.
\textbf{(b)} Upper panel: experimental adsorption
isotherm of CO$_2$ in MIL-53~(Al) at 304~K,\cite{MIL53_JACS} as well as
Langmuir fits of the 0--5~bar and 9--30~bar regions (in red and
blue, respectively). Lower panel: osmotic thermodynamic potential as a
function of CO$_2$ pressure for the \textbf{lp} (in blue) and \textbf{np}
(in red) structures of MIL-53~(Al), calculated from the Langmuir fits.}
\end{center} \end{figure}

The metal--organic framework MIL-53 has attracted a lot of attention due
to the massive flexibility it exhibits, including structural
characterization,\cite{Serre, MIL53_AdvMater} adsorption of strategic
gases H$_2$, CO$_2$ and CH$_4$\cite{MIL53_JACS, MIL53_H2} and
simulations.\cite{Ramsahye2007, Ramsahye_JPCC, Coombes} The MIL-53
framework topology is formed of unidimensional chains of corner-sharing
MO$_4$(OH)$_2$ octahedra (M$=$Al$^{3+}$, Cr$^{3+}$) linked by
1,4-benzenedicarboxylate (BDC) ligands, which results in linear
lozenge-shaped channels large enough to accommodate small guest
molecules. This structure may oscillate between two distinct states, a
large pore form (\textbf{lp}) and a narrow pore form (\textbf{np}), upon
adsorption and desorption of gases; there is a $\sim 38$\% difference in
cell volume between these two forms. Both structures are depicted in
Figure~\ref{fig:iso_CO2_MIL53}a.

The adsorption isotherm of CO$_2$ in MIL-53 at 304~K exhibit a step at
approximately 6~bar (upper panel of Figure~\ref{fig:iso_CO2_MIL53}b),
demonstrated to emanate from a structural transition upon adsorption from
the CO$_2$-loaded \textbf{np} form to the CO$_2$-loaded \textbf{lp}
one.\cite{MIL53_AdvMater} Moreover, as the most favorable guest-free
MIL-53 is the large pore form at room temperature, another transition,
\textbf{lp}$\rightarrow$\textbf{np}, is expected at low pressure.
Although it was not seen in the original adsorption
isotherm,\cite{MIL53_JACS} this low pressure transition was very recently
confirmed by a combined simulation and microcalorimetry
study.\cite{Coombes}

Recently, a lot of effort has been put into understanding the energetics
of these two guest-induced structural transitions, by Density Functional
Theory\cite{Ramsahye_JPCC} and forcefield-based calculations using
rigid\cite{Ramsahye2007} or flexible\cite{Coombes} MIL-53 structures.
Here, we bring the discussion one step further by giving valuable insight
on the thermodynamics of this guest-induced structural transition using
our method and the simple model developed above, yielding quantitative
information on the relative stability of the two MIL-53 structures, which
are virtually impenetrable experimentally.

The upper pannel of Figure~\ref{fig:iso_CO2_MIL53}b shows the experimental
adsorption isotherm of CO$_2$ in MIL-53~(Al). The step between 5~and
9~bar is a signature of the transition between the two phases of
MIL-53~(Al), separating the part of the isotherm that corresponds to the
\textbf{np} structure ($P < 5$~bar) and the part where the structure is
fully open ($P > 9$~bar). Both parts were satisfactorily fitted by
Langmuir isotherms (Fig.~\ref{fig:iso_CO2_MIL53}b) and Henry constants
were extracted (they are the slopes of the Langmuir fits at $P=0$). The
latter show clearly that the adsorption affinity of CO$_2$ is much higher
in the \textbf{np} structure than in the \textbf{lp} one (3.5~times
higher; $K\e{\textbf{np}} \simeq 9.0\times
10^{-5}$~mol\,kg$^{-1}$\,Pa$^{-1}$ and $K\e{\textbf{lp}} \simeq 2.6\times
10^{-5}$~mol\,kg$^{-1}$\,Pa$^{-1}$), in line with the energetics
published so far.\cite{MIL53_JACS, Ramsahye_JPCC, Coombes} Indeed, with
such features, MIL-53 exactly fits in \emph{case~c} of our taxonomy: the
\textbf{lp} form is more stable in the absence of adsorbate and has a
larger pore volume, but has a smaller affinity for CO$_2$ than
\textbf{np}. Thus, according to our model, MIL-53 is expected to exhibit
a double structural transition upon gas adsorption if the affinity is
high enough: when empty and at very low pressure, the \textbf{lp}
structure is intrinsically more stable; when pressure increases, the high
affinity of the \textbf{np} structure leads to a
\textbf{lp}$\rightarrow$\textbf{np} transition; upon further rise of gas
pressure, the larger pore volume of the open structure makes it favorable
again and a \textbf{np}$\rightarrow$\textbf{lp} transition should be
observed. Our simple model predicts that, for this type~I adsorption of
gas in MIL-53 (Al, Cr), there is either no structural transition or two
of them, depending on the nature of the gas. In the case of CO$_2$, the
experimental observation of one transition leads to the conclusion that
another one must exist, at lower pressure, which was indeed recently
evidenced.\cite{Coombes} By contrast, CH$_4$ has a much smaller
adsorption affinity in MIL-53 and appears not to induce any structural
transition,\cite{MIL53_JACS} in agreement with our discussion above.
These results highlight the correctness of the model and its ability to
predict the occurence or the absence of structural transitions for a
given adsorbate, on the sole basis of its respective affinities for the
host structures.

To quantify the effects discussed above, we have calculated the
thermodynamic osmotic potential, $\Omega\e{os}(P)$, of the \textbf{lp}
and \textbf{np} phases of MIL-53~(Al) as a function of CO$_2$ pressure
using the Eq.~\ref{eq:Omega_os} and the fitted isotherms described above.
From the experimental isotherm, we assume the equilibrium pressure for
the \textbf{np}$\rightarrow$\textbf{lp} transition to be around 5~bar. We
then find a difference in free energy between the two structures of
$\Delta F\e{\textbf{lp}-\textbf{np}} \simeq 2.5$~kJ/mol per unit cell. A
similar value was obtained for the chromium-containing form of MIL-53.
The osmotic potential profiles (Fig.~\ref{fig:iso_CO2_MIL53}b, lower
panel) clearly confirm the existence of two structural transitions upon
adsorption and predict the low pressure
\textbf{lp}$\rightarrow$\textbf{np} transition to happen at 0.3~bar. It
is noteworthy that this predicted value of the transition pressure is in
very good agreement with the experimental value of 0.25~bar found very
recently by microcalorimetry,\cite{Coombes} which is once again
indicative of the predictive quality of the method exposed here. It also
helps explaining why this low pressure transition was not visible on the
original adsorption isotherm.


\section*{Conclusions}

This article focuses on the specific class of hybrid materials exhibiting
guest-induced structural transitions upon gas adsorption. Studies in this
rapidly growing field mainly revolve around the structural
characterization of these systems, the determination of their adsorption
properties and the microscopic understanding of the coupling between the
framework and adsorbate, e.g. by atomistic simulations. Here, we expose a
general thermodynamic framework for gas adsorption in flexible hybrid
materials and put forward a taxonomy of guest-induced structural
transitions. This allows us to predict the occurence or the absence of
transitions of the host on the sole basis of a few key parameters: pore
volumes, adsorption affinities and relative free energies of the
framework structures involved in the transitions. Conversely, we show
that these relative free energies can be extracted from experimental
stepped isotherms, therefore quantifying the energetics of the phase
transitions, which are especially difficult to access experimentally,
however vital to the fundamental understanding of these systems. The
robustness of our thermodynamic approach is illustrated through the
analysis of three distinct cases of guest-induced transitions: the
hysteretic H$_2$ adsorption in a cobalt-based 3D framework, the
gate-opening process in an interdigitated coordination polymer and the
guest-dependent ``breathing'' phenomenon of MIL-53. We show that the free
energy differences between the host structures involved in these
guest-induced transitions fall into the 2--6~kJ/mol range. To our
knowledge, this is the first general thermodynamic framework successfully
rationalizing the variety of behaviors reported for this class of
materials. We believe this work should prove very useful to explore new
guest-induced transitions and further complement the current experimental
and simulation approaches in the field.


\section*{Acknowledgments}

This work was supported by the EPSRC (Advanced Research Fellowship
awarded to CMD) and the FP-6 -- European funding STREP ``DeSanns''
(SES6-CT-2005-020133) and ``SURMOF'' (NMP4-CT-2006-032109).

\section*{Supporting Information Available} Calculation details for the
derivation of the taxonomy (Appendix~A).

\clearpage

\bibliography{article}

\providecommand{\url}[1]{\texttt{#1}}
\providecommand{\refin}[1]{\\ \textbf{Referenced in:} #1}
\begin{thebibliography}{10}

\bibitem{Kitagawa_rev1}
Kitagawa,~S.; Kitaura,~R.; Noro,~S. \emph{Angew. Chem. Int. Ed.} \textbf{2004},
  \emph{43}, 2334--2375.

\bibitem{Rosseinsky_rev}
Rosseinsky,~M.~J. \emph{Microporous Mesoporous Mater.} \textbf{2004},
  \emph{73}, 15--30.

\bibitem{Cheetham2006}
Cheetham,~A.~K.; Rao,~C. N.~R.; Feller,~R.~K. \emph{Chem. Commun.}
  \textbf{2006},  4780--4795.

\bibitem{Cheetham2008}
Rao,~C. N.~R.; Cheetham,~A.~K.; Thirumurugan,~A. \emph{Journal of Physics:
  Condensed Matter} \textbf{2008}, \emph{20}, 083202.

\bibitem{Ferey_AccChemRes}
F{\'e}rey,~G.; Mellot-Draznieks,~C.; Serre,~C.; Millange,~F. \emph{Acc. Chem.
  Res.} \textbf{2005}, \emph{38}, 217--225.

\bibitem{Kepert}
Kepert,~C.~K. \emph{Chem. Commun.} \textbf{2006},  695--700.

\bibitem{Rowsell}
Rowsell,~J. L.~C.; Millward,~A.~R.; Park,~K.~S.; Yaghi,~O.~M. \emph{J. Am.
  Chem. Soc.} \textbf{2004}, \emph{126}, 5666--5667.

\bibitem{Latroche}
Latroche,~M.; Surbl{\'e},~S.; Serre,~C.; Mellot-Draznieks,~C.;
  Llewellyn,~P.~L.; Lee,~J.-H.; Chang,~J.-S.; Jhung,~S.~H.; F{\'e}rey,~G.
  \emph{Angew. Chem. Int. Ed.} \textbf{2006}, \emph{45}, 8227--8231.

\bibitem{Eddaoudi}
Eddaoudi,~M.; Kim,~J.; Rosi,~N.; Vodak,~D.; Wachter,~J.; {O'Keeffe},~M.;
  Yaghi,~O.~M. \emph{Science} \textbf{2002}, \emph{295}, 469--472.

\bibitem{Banerjee}
Banerjee,~R.; Phan,~A.; Wang,~B.; Knobler,~C.; Furukawa,~H.; {O'Keeffe},~M.;
  Yaghi,~O.~M. \emph{Science} \textbf{2008}, \emph{319}, 939--943.

\bibitem{Bradshaw}
Bradshaw,~D.; Claridge,~J.~B.; Cussen,~E.~J.; Prior,~T.~J.; Rosseinsky,~M.~J.
  \emph{Acc. Chem. Res.} \textbf{2005}, \emph{38}, 273--282.

\bibitem{Rosseinsky}
Fletcher,~A.~J.; Thomas,~K.~M.; Rosseinsky,~M.~J. \emph{J. Solid State Chem.}
  \textbf{2005}, \emph{178}, 2491--2510.

\bibitem{Kitagawa_rev}
Kitagawa,~S.; Uemura,~K. \emph{Chem. Soc. Rev.} \textbf{2005}, \emph{34},
  109--119.

\bibitem{Kitagawa_rev2}
Uemura,~K.; Matsuda,~R.; Kitagawa,~S. \emph{J. Solid State Chem.}
  \textbf{2005}, \emph{178}, 2420--2429.

\bibitem{Kondo}
Kondo,~A.; Noguchi,~H.; Carlucci,~L.; Proserpio,~D.~M.; Ciani,~G.; Kajiro,~H.;
  Ohba,~T.; Kanoh,~H.; Kaneko,~K. \emph{J. Am. Chem. Soc.} \textbf{2007},
  \emph{129}, 12362--12362.

\bibitem{Noguchi}
Noguchi,~H.; Kondo,~A.; Hattori,~Y.; Kajiro,~H.; Kanoh,~H.; Kaneko,~K. \emph{J.
  Phys. Chem. C} \textbf{2007}, \emph{111}, 248--254.

\bibitem{Maji}
Maji,~T.~K.; Mostafa,~G.; Matsuda,~R.; Kitagawa,~S. \emph{J. Am. Chem. Soc.}
  \textbf{2005}, \emph{127}, 17152--17153.

\bibitem{Chen}
Chen,~B.; Ma,~S.; Hurtado,~E.~J.; Lobkovsky,~E.~B.; Liang,~C.; Zhu,~H.; Dai,~S.
  \emph{J. Phys. Chem. B} \textbf{2007}, \emph{111}, 6101--6103.

\bibitem{Shimomura}
Shimomura,~S.; Horike,~S.; Matsuda,~R.; Kitagawa,~S. \emph{J. Am. Chem. Soc.}
  \textbf{2007}, \emph{129}, 10990--10990.

\bibitem{Matsuda}
Matsuda,~R.; Kitaura,~R.; Kitagawa,~S.; Kubota,~Y.; Belosludov,~R.~V.;
  Kobayashi,~T.~C.; Sakamoto,~H.; Chiba,~T.; Takata,~M.; Kawazoe,~Y.; Mita,~Y.
  \emph{Nature} \textbf{2005}, \emph{436}, 238--241.

\bibitem{Goto}
Goto,~M.; Furukawa,~M.; Miyamoto,~J.; Kanoh,~H.; Kaneko,~K. \emph{Langmuir}
  \textbf{2007}, \emph{23}, 5264--5266.

\bibitem{Hu}
Hu,~S.; Zhang,~J.-P.; Li,~H.-X.; Tong,~M.-L.; Chen,~X.-M.; Kitagawa,~S.
  \emph{Cryst. Growth Des.} \textbf{2007}, \emph{7}, 2286--2289.

\bibitem{Tanaka}
Tanaka,~D.; Nakagawa,~K.; Higuchi,~M.; Horike,~S.; Kubota,~Y.;
  Kobayashi,~T.~C.; Takata,~M.; Kitagawa,~S. \emph{Angew. Chem. Int. Ed.}
  \textbf{2008}, \emph{47}, 3914--3918.

\bibitem{Long}
Choi,~H.~J.; Dinc{\-a},~M.; Long,~J.~R. \emph{J. Am. Chem. Soc.} \textbf{2008},
  \emph{130}, 7848--7850.

\bibitem{IUPAC_isotherms}
Sing,~K. S.~W.; Everett,~D.~H.; Haul,~R.~A.; Moscou,~L.; Pierotti,~R.~A.;
  Rouquerol,~J.; Siemieniewska,~T. \emph{Pure Appl. Chem.} \textbf{1985},
  \emph{57}, 603--619.

\bibitem{Millward}
Millward,~A.~R.; Yaghi,~O.~M. \emph{J. Am. Chem. Soc.} \textbf{2005},
  \emph{127}, 17998--17999.

\bibitem{Walton}
Walton,~K.~S.; Millward,~A.~R.; Dubbeldam,~D.; Frost,~H.; Low,~J.~J.;
  Yaghi,~O.~M.; Snurr,~R.~Q. \emph{J. Am. Chem. Soc.} \textbf{2008},
  \emph{130}, 406--407.

\bibitem{Smit}
Smit,~B.; Maesen,~T. L.~M. \emph{Nature} \textbf{2002}, \emph{374}, 42--44.

\bibitem{Lachet}
Lachet,~V.; Boutin,~A.; Pellenq,~R. J.-M.; Nicholson,~D.; Fuchs,~A.~H. \emph{J.
  Phys. Chem.} \textbf{1996}, \emph{100}, 9006--9013.

\bibitem{Kaneko}
Li,~D.; Kaneko,~K. \emph{Chem. Phys. Lett.} \textbf{2001}, \emph{335}, 50--56.

\bibitem{Kitaura}
Kitaura,~R.; Seki,~K.; Akiyama,~G.; Kitagawa,~S. \emph{Angew. Chem. Int. Ed.}
  \textbf{2003}, \emph{42}, 428--431.

\bibitem{Serre}
Serre,~C.; Millange,~F.; Thouvenot,~C.; Nogues,~M.; Marsolier,~G.;
  Lou{\"e}r,~D.; F{\'e}rey,~G. \emph{J. Am. Chem. Soc.} \textbf{2002},
  \emph{124}, 13519--13526.

\bibitem{MIL88}
Mellot-Draznieks,~C.; Serre,~C.; Surbl{\'e},~S.; Audebrand,~N.; F{\'e}rey,~G.
  \emph{J. Am. Chem. Soc.} \textbf{2005}, \emph{127}, 16273--16278.

\bibitem{MIL88_2}
Serre,~C.; Mellot-Draznieks,~C.; Surbl{\'e},~S.; Audebrand,~N.; Filinchuk,~Y.;
  F{\'e}rey,~G. \emph{Science} \textbf{2007}, \emph{315}, 1828--1831.

\bibitem{MIL53_AdvMater}
Serre,~C.; Bourrelly,~S.; Vimont,~A.; Ramsahye,~N.~A.; Maurin,~G.;
  Llewellyn,~P.~L.; Daturi,~M.; Filinchuk,~Y.; Leynaud,~O.; Barnes,~P.; G.,~F.
  \emph{Adv. Mater.} \textbf{2007}, \emph{19}, 2246--2251.

\bibitem{Coombes}
Coombes,~D.~S.; Bell,~R.; Bourrelly,~S.; Llewellyn,~P.~L.;
  Mellot-Draznieks,~C., submitted for publication.

\bibitem{Ramsahye_JPCC}
Ramsahye,~N.~A.; Maurin,~G.; Bourrelly,~S.; Llewellyn,~P.~L.; Serre,~C.;
  Loiseau,~T.; Devic,~T.; F{\'e}rey,~G. \emph{J. Phys. Chem. C} \textbf{2008},
  \emph{112}, 514--520.

\bibitem{Brennan}
Brennan,~J.~K.; Madden,~W.~G. \emph{Macromolecules} \textbf{2002}, \emph{35},
  2827--2834.

\bibitem{dePablo_osmotic}
Banaszak,~B.~J.; Faller,~R.; de~Pablo,~J.~J. \emph{J. Chem. Phys.}
  \textbf{2004}, \emph{120}, 11304--11315.

\bibitem{jeffroy}
Jeffroy,~M.; Fuchs,~A.~H.; Boutin,~A. \emph{Chem. Commun.} \textbf{2008},
  3275--3277.

\bibitem{Snurr}
Snurr,~R.~Q.; Bell,~A.~T.; Theodorou,~D.~N. \emph{J. Phys. Chem.}
  \textbf{1994}, \emph{98}, 5111--5119.

\bibitem{Shen}
Shen,~J.; Monson,~P.~A. \emph{Mol. Phys.} \textbf{2002}, \emph{100},
  2031--2039.

\bibitem{Faure}
Faure,~F.; Rousseau,~B.; Lachet,~V.; Ungerer,~P. \emph{Fluid Phase Equilibria}
  \textbf{2007}, \emph{261}, 168--175.

\bibitem{Peterson}
Peterson,~B.~K.; Gubbins,~K.~E. \emph{Mol. Phys.} \textbf{1987}, \emph{62},
  215--226.

\bibitem{Puibasset}
Puibasset,~J.; Pellenq,~R. J.~M. \emph{J. Chem. Phys.} \textbf{2005},
  \emph{122}, 094704.

\bibitem{Cailliez}
Cailliez,~F.; Stirnemann,~G.; Boutin,~A.; Demachy,~I.; Fuchs,~A.~H. \emph{J.
  Phys. Chem. C} \textbf{2008}, \emph{112}, 10435--10445.

\bibitem{Navrotsky}
Navrotsky,~A. \emph{Phys. Chem. Miner.} \textbf{1997}, \emph{24}, 222--241.

\bibitem{note_fit_type}
{The isotherm presented as illustration of the method has two parts that are
  each isotherms of type~I, and thus appropriately fitted by Langmuir
  isotherms, but the method presented here is more general and can work as long
  as you have some \emph{a priori} idea of the form of the isotherm and an
  adequate fitting function.}

\bibitem{note_density}
{We assume that the fluid density in the pores at sufficiently high pressure is
  identical for all host phases, but the phenomenological discussion remains
  the same without this assumption, except that the pore volumes,
  $V\e{p}\ex{($i$)}$, are replaced by the saturation values of the isotherms,
  $N\e{max}^i$.}

\bibitem{note_case_a}
{It is to be noted that, while the existence of this single transition is
  independent of the values of $K_1$, $K_2$ and $\Delta F\e{host}$, the
  pressure at which it occurs is not: the smaller $\Delta F\e{host}$ or the
  larger $\Delta K$, the lower the transition pressure.}

\bibitem{MIL53_JACS}
Bourrelly,~S.; Llewellyn,~P.~L.; Serre,~C.; Millange,~F.; Loiseau,~T.;
  F{\'e}rey,~G. \emph{J. Am. Chem. Soc.} \textbf{2005}, \emph{127},
  13519--13521.

\bibitem{Collins}
Collins,~D.~J.; Zhou,~H.-C. \emph{J. Mater. Chem.} \textbf{2007}, \emph{17},
  3154--3160.

\bibitem{Zhao}
Zhao,~X.; Xiao,~B.; Fletcher,~A.~J.; Thomas,~K.~M.; Bradshaw,~D.;
  Rosseinsky,~M.~J. \emph{Science} \textbf{2004}, \emph{306}, 1012--1015.

\bibitem{Omary}
Yang,~C.; Wang,~X.; Omary,~M.~A. \emph{J. Am. Chem. Soc} \textbf{2007},
  \emph{129}, 15454--15455.

\bibitem{note_pv}
{ We need to make this approximation because the unit cell parameters of
  structure~\textbf{A1} are not known. It is justified by the fact that, as
  mentionned earlier, these terms have little influence in the range of
  pressure considered here. }.

\bibitem{MIL53_H2}
F{\'e}rey,~G.; Latroche,~M.; Serre,~C.; Millange,~F.; Loiseau,~T.;
  Percheron-Guegan,~A. \emph{Chem. Commun.} \textbf{2003},  2976--2977.

\bibitem{Ramsahye2007}
Ramsahye,~N.~A.; Maurin,~G.; Bourrelly,~S.; Llewellyn,~P.~L.; Loiseau,~T.;
  Serre,~C.; F{\'e}rey,~G. \emph{Chem. Comm.} \textbf{2007},  3261--3263.

\end{thebibliography}

\clearpage

\vspace*{2cm}

\begin{center}
\textbf{\LARGE Table of Contents Graphic}
\end{center}

\vspace*{1cm}

\begin{center}


\includegraphics[height=3.5cm]{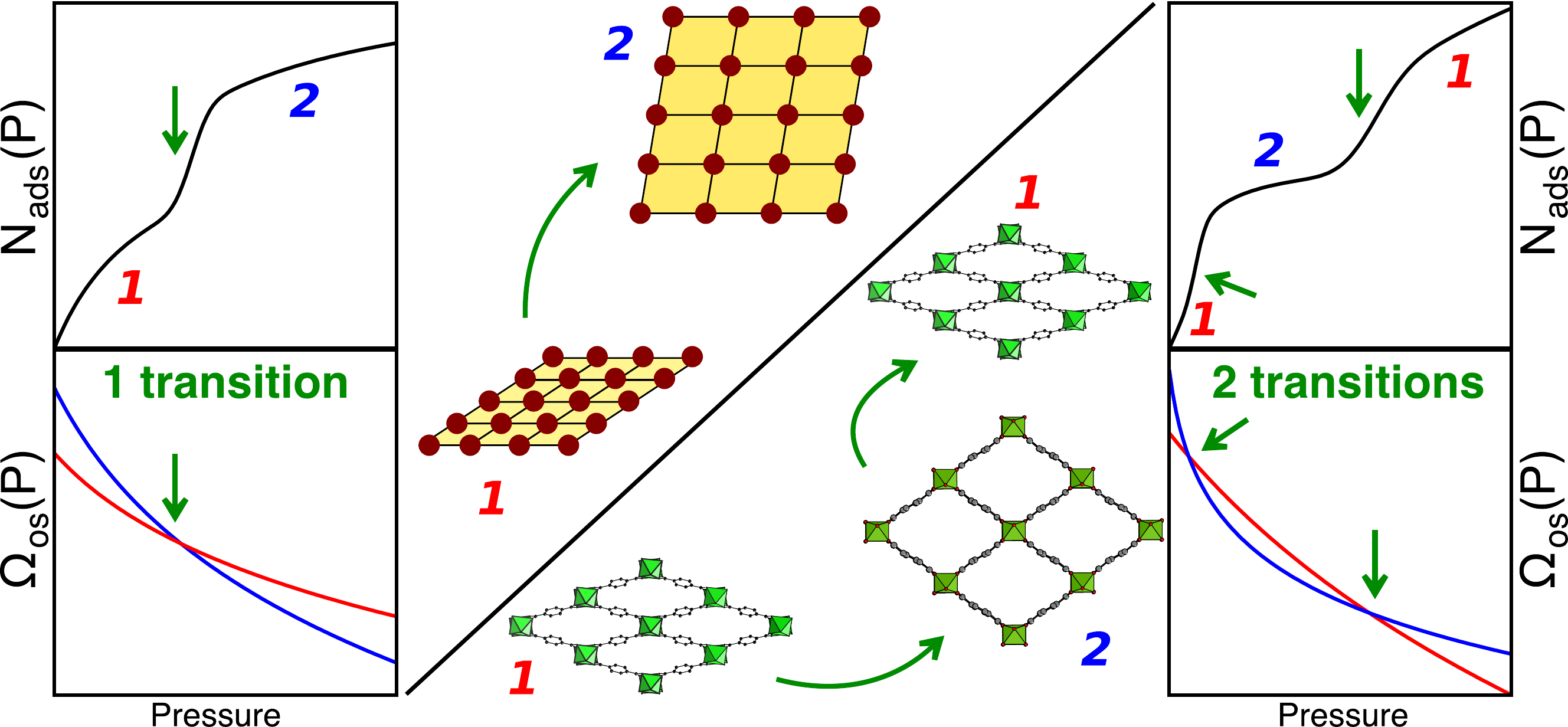}

\end{center}

\clearpage
\setcounter{page}{1}
\renewcommand\thepage{S\arabic{page}}

\begin{center}

{\Large Supplementary information for:}

\vskip 1cm

{\Large Thermodynamics of guest-induced structural transitions in hybrid
organic--inorganic frameworks}

\vskip 5mm

{\large F.-X. Coudert, M. Jeffroy,
A.H. Fuchs, A. Boutin and C. Mellot-Draznieks}

\end{center}

\section*{Appendix~A: Calculation details}

Starting with the expression of $\Delta\Omega\e{os}(P)$ in
Equation~\ref{eq:diff_omega_langmuir}, we want to study the sign of
$\Delta\Omega\e{os}(P)$ and the solutions of the equation
$\Delta\Omega\e{os}(P) = 0$. To do so, we first compute the derivative of
$\Delta\Omega\e{os}(P)$ and we can show that: \[ \frac{\mathrm{d}
\Delta\Omega\e{os}}{\mathrm{d} P} = -\frac{\rho RT}{ \left(\rho
V\e{p}\ex{(1)} + K_1 P\right) \left(\rho V\e{p}\ex{(2)} + K_2 P\right) }
\left( K_1K_2 \Delta V\e{p} P + \rho V\e{p}\ex{(1)} V\e{p}\ex{(2)} \Delta
K \right) \] The sign of the derivative is thus the same as that of
$-\left(K_1K_2 \Delta V\e{p} P + \rho V\e{p}\ex{(1)} V\e{p}\ex{(2)}
\Delta K\right)$. Four different cases need to be considered, depending
on the sign of both $\Delta K$ and $\Delta V\e{p}$: \begin{itemize}
\item If $\Delta V\e{p} > 0$ and $\Delta K > 0$, the derivative will
always we always be negative, and $\Delta\Omega\e{os}(P)$ is strictly
decreasing. As $\Delta\Omega\e{os}(P=0) = \Delta F\e{host} > 0$, the
equation $\Delta\Omega\e{os}(P) = 0$ has one unique solution.
\item If $\Delta V\e{p} > 0$ and $\Delta K < 0$, the derivative will be
positive around $P=0$ and will become negative at higher pressure.
$\Delta\Omega\e{os}(P)$ has a positive value at $P=0$, is first
increasing, then decreasing; the equation $\Delta\Omega\e{os}(P) = 0$
will once again have one solution (and only one).
\item If $\Delta V\e{p} < 0$ and $\Delta K < 0$, the derivative will
always be positive and $\Delta\Omega\e{os}(P)$ is increasing: at its
value in $P=0$ is positive, the equation $\Delta\Omega\e{os}(P) = 0$ will
have no solution.
\item If $\Delta V\e{p} < 0$ and $\Delta K > 0$, the derivative will be
negative at small~$P$ and positive at larger~$P$. Starting with a
positive value at $P=0$, $\Delta\Omega\e{os}(P)$ will thus first decrease
and then increase. Depending on the value it reaches at its minimum, the
equation $\Delta\Omega\e{os}(P) = 0$ will have zero, one (in the limiting
case) or two solutions.
\end{itemize} It is worth noting that the first two cases can be joined
into a single case: if $\Delta V\e{p} > 0$, the equation
$\Delta\Omega\e{os}(P) = 0$ has one unique solution.

\end{document}